\documentclass[conference]{IEEEtran}

\usepackage{graphicx}
\usepackage{cite}
\usepackage{amsfonts}
\graphicspath{{eps/}} \DeclareGraphicsExtensions{.eps}

\usepackage[cmex10]{amsmath}
\usepackage{amssymb}
 \IEEEoverridecommandlockouts
\ifCLASSINFOpdf
\else
\fi
\begin{document}

%
%
\title{An Energy Efficient Semi-static Power Control and Link Adaptation Scheme in UMTS HSDPA
\thanks{Corresponding author: Ling Qiu,lqiu@ustc.edu.cn.}\thanks{This paper is accepted in EURASIP Journal on Wireless Communications and Networking, special issue on Green Radio}}

\author{\IEEEauthorblockN{Yi Huang, Jie Xu, Ling Qiu}
\IEEEauthorblockA{
University of Science and Technology of China (USTC),\\
Hefei, China, 230027\\
Email: yihuang@mail.ustc.edu.cn, suming@mail.ustc.edu.cn, lqiu@ustc.edu.cn}}


%


\maketitle


\begin{abstract}
High speed downlink packet access (HSDPA) has been successfully applied in commercial systems and improves user experience significantly. However, it incurs substantial energy consumption. In this paper, we address this issue by proposing a novel energy efficient semi-static power control and link adaptation scheme in HSDPA. Through estimating the EE under different modulation and coding schemes (MCSs) and corresponding transmit power, the proposed scheme can determine the most energy efficient MCS level and transmit power at the Node B. And then the Node B configure the optimal MCS level and transmit power. In order to decrease the signaling overhead caused by the configuration, a dual trigger mechanism is employed. After that, we extend the proposed scheme to the multiple input multiple output (MIMO) scenarios. Simulation results confirm the significant EE improvement of our proposed scheme. Finally, we give a discussion on the potential EE gain and challenge of the energy efficient mode switching between single input multiple output (SIMO) and MIMO configuration in HSDPA.
\end{abstract}

\section*{Keywords}
HSDPA, energy efficiency, power control, link adaptation.


\section{Introduction}
How to acquire higher throughput with lower power consumption has become an important challenge for the future wireless communication systems \cite{refer1}. ``Moore's Law'' renders the use of ever more powerful Information and communications technology (ICT) systems for the mass market. In order to transport this exponentially rising amount of available data to the user in an acceptable time, the transmission rate in cellular network rises at the speed of nearly 10 times every 5 years, meanwhile the energy consumption doubles every 5 years, as illustrated in \cite{refer2}.

High speed downlink packet access (HSDPA) has been successfully applied commercially, which brings high spectral efficiency (SE) and enhances user experience. According to \cite{refer3}, HSDPA has introduced a new downlink physical channel called  high speed physical downlink shared channel (HS-PDSCH), and some new features such as adaptive modulation and coding scheme (AMC), hybrid automatic repeat request (HARQ), fast scheduling and multiple input multiple output (MIMO). Thus it improves the downlink peak data rate and system throughput greatly. For the MIMO technology in HSDPA, the so-called dual stream transmit adaptive antennas (D-TxAA) is applied, in which the Node B would select single stream mode or dual stream mode based on the channel conditions.

To the best of the authors' knowledge, most of the previous research works focused on spectral efficient schemes in UMTS HSDPA and only a few
literatures focused on the network energy savings \cite{refer4,refer5}. In \cite{refer4}, the authors proposed to switch off a second carrier
in Dual-Cell HSDPA to save energy through exploiting the network traffic variations. And the authors in \cite{refer5} investigated the possibility
of cutting down the energy consumption of the wireless networks by reducing the number of active cells when the traffic load is low. These works
mainly considered energy savings from a network point of view. However, there is no literature focusing on the link level energy efficient
schemes in HSDPA, which is also an important aspect in green communication research.

Energy efficiency (EE) is always defined as the transmission rate divided by the total power consumption, which represents the number of
information bits transmitted over unit energy consumption measured in bits/Joule. In the previous works considering EE from a link level
perspective \cite{refer6,refer7,refer8,refer9,refer10}, EE maximization problems are formulated and solved based on Shannon capacity, in
which the impact of constant circuit power is involved. It is demonstrated that joint power control and link adaptation is an effective method to improve the EE. However, practical modulation and channel coding schemes are not considered in these works and the users' quality of service (QoS) constraints are not taken into account either. Moreover, as the fast power control is not available in HS-PDSCH due to the functionality of AMC and HARQ, it is hard to apply joint power control and link adaptation in the HSDPA system directly.

In this paper, we will discuss the potential link level energy saving in HSDPA. First, a power model including dynamic circuit power related with antenna number is taken into account. Based on this model, we propose a practical semi-static joint power control and link adaptation method to improve EE, while guaranteeing the users' transmission rate constraints.  As fast power control is no longer supported, we propose a dual
trigger mechanism to perform the method semi-statically. After that, we extend the scheme to the MIMO HSDPA systems. Simulation results confirm the significant EE improvement of our proposed method. Finally, we give a discussion on the potential EE gain and challenges of the energy efficient mode switching between single input multiple output (SIMO) and MIMO configuration.

The rest of the paper is organized as follows. Section 2 introduces the preliminaries. Section 3 proposes the energy efficient power control and link adaptation scheme in the single input single output (SISO) HSDPA systems. The extension of the scheme to the MIMO HSDPA systems is presented in Section 4. Simulation results and discussion are given in Section 5, and finally Section 6 concludes this paper.

\section{Preliminaries}

In this section, preliminaries are provided. The system model and power model are introduced at first. The theoretic SE-EE tradeoff is then provided to help the description.
\subsection{System Model}

We consider the system with a single Node B and a single user in this paper, but note that our work can be extended to the multi-user scenario
easily. We assume that the Node B has a maximum transmit power constraint $P_{\max}$ and the user has a minimum modulation and coding scheme (MCS)
constraint $\theta_{\min}$ which can be viewed as the QoS requirements.

\subsubsection{Link Adaptation Scheme and D-TxAA Functionality in HSDPA}
The traditional link adaptation of the HSDPA systems is illustrated as follows. First, the Node B determines the transmit power of HS-PDSCH.
Once the transmit power is determined, it cannot be changed frequently, due to the existence of AMC and HARQ. The user measures the channel quality
between the Node B and itself and feeds back a channel quality indication (CQI) to the Node B. The feedback CQI corresponds to a MCS level which is
always chosen to maximize the transmission rate under a certain bit error rate (BER). Then the Node B delivers data to the user with the MCS level. In this way, the transmission parameters can be adjusted according to current channel conditions and thus high throughput can be provided.

D-TxAA is selected as the MIMO scheme for HSDPA in 3GPP specification Release 7 \cite{refer11}. Two antennas at the Node B and the user are
supported. Specifically, the Node B sends buffered data through either one or two independent data streams at the physical layer. At first, the
user determines the preferred CQI for the single stream mode and the preferred pair of CQIs for the dual stream mode. After comparing the
transmission rates of the two modes, the user can choose the better mode and corresponding CQI(s) and then feed them back to the Node B. Thus, the Node B can decide the mode and corresponding MCS level(s).

In addition to CQI feedback, the user also reports precoding control indicator (PCI) index which indicates the optimal precoding weights $\{w_1,w_2\}$
for the primary stream, based on which precoding weights $\{w_3,w_4\}$ for the second stream can be calculated. The precoding weights are
defined as follows\cite{refer11}:
\begin{equation} \label{eq1}
\begin{array}{l}
 w_1  = w_3  = 1/\sqrt 2 , \\
 w_4  =  - w_2 , \\
 w_2  = \displaystyle [\frac{{1 + j}}{2},\frac{{1 - j}}{2},\frac{{ - 1 + j}}{2},\frac{{ - 1 - j}}{2}]. \\
\end{array}
\end{equation}

\subsection{Power Consumption Model}
Power consumption model here is based on \cite{refer10} in order to capture the effect of transmit antenna number. Denote the number of active transmit antennas as $M_{\rm a}$ and transmit power as $P$. The total power consumption of Node B is divided into three parts. The first part is
the power conversion (PC) power
\begin{equation} \label{eq2}
\begin{array}{l}
P_{{\rm{PC}}}  = \frac{\displaystyle P}{\displaystyle \eta },
\end{array}
\end{equation}
accounting for the power consumption in the power amplifier and related feeder loss, in which $\eta$ is the PC efficiency. The second part is
the dynamic circuit power which corresponds to antenna number $M_{\rm a}$ and can be given by:
\begin{equation} \label{eq3}
\begin{array}{l}
P_{\rm Dyn}  = M_{\rm{a}} P_{\rm{cir}},
\end{array}
\end{equation}
representing circuit power consumption for Radio Frequency(RF) and signal processing. The third part is the static power $P_{\rm Sta}$
related to cooling loss, battery backup and power supply loss, which is independent of $M_{\rm a}$ and $P_{\rm PC}$. The total power consumption can be modeled as
\begin{equation} \label{eq4}
\begin{array}{l}
P_{{\rm{total}}}  = P_{{\rm{PC}}}  + P_{{\rm{Dyn}}}  + P_{{\rm{Sta}}}.
\end{array}
\end{equation}

\subsection{SE and EE Trade-off}
Before introducing our proposal, we need to have a discussion about the theoretical basis of the energy efficient power control and link
adaptation scheme. According to the Shannon capacity, SE and EE of a SISO additive white Gaussian noise (AWGN) channel can be expressed as
\begin{equation} \label{eq5}
\begin{array}{l}
\mu = \displaystyle \log _2 (1 + \frac{P}{{N_0 W}})
\end{array}
\end{equation}
and
\begin{equation} \label{eq6}
\begin{array}{l}
\xi  = \displaystyle W\log _2 (1 + \frac{P}{{N_0 W}})/{P_{\rm{total}}}
\end{array}
\end{equation}
respectively, where $W$ and $N_0$ represent system bandwidth and the noise density respectively.

It is obvious from (\ref{eq5}) that the transmit power $P$ is exponentially increasing as a function of the SE with the assumption of constant bandwidth and noise power. In other words, higher SE incurs significant increase of energy consumption. In fact, EE is monotonically decreasing with SE if only the transmit power is
considered\cite{refer12}. Thus in order to improve EE, Node B should reduce the transmit power. However, the existence of
practical $P_{{\rm{Dyn}}}$ and $P_{{\rm{Sta}}}$ breaks the monotonic relation between SE and EE, so balancing the $P_{\rm{PC}}$,
$P_{{\rm{Dyn}}}$ and $P_{{\rm{Sta}}}$ is also important to increase EE. Figure 1 shows the EE-power and SE-power relations in an AWGN channel with the theoretical Shannon capacity formula. As indicated in Figure 1, there exists a globally optimal transmit power for EE. Moreover, based on the Shannon
capacity, we can obtain the explicit close-form solution of the globally optimal EE and optimal transmit power, and some examples in MIMO systems
can be found in \cite{refer10}.

However, one may argue that whether the relation between EE and SE still satisfies in the HSDPA systems when practical AMC and HARQ are taken into account. Fortunately, we confirm this principle through the HSDPA link level simulation and the result with SISO channels based on TABLE G is shown in Figure 2. The MIMO systems with D-TxAA have the similar relations, which is shown later in this paper. Although this trend is still fulfilled, the challenge in the HSDPA systems is that the explicit close-form solution to obtain the optimal transmit power and corresponding MCS level is no longer available when practical AMC and HARQ are applied here.  To meet this challenge, we will solve this problem through a novel EE estimation mechanism in the rest of this paper. Besides, the data rate constraints are considered due to the users' QoS requirements in practice. According to the constraints, we should find the feasible transmit power region first, and then determine the transmit power with constrained optimal EE based on the feasible region. More details will be given in the next section.

\section{Energy Efficient Power Control and Link Adaptation Scheme in SISO Systems}

A semi-static power control and link adaptation method is proposed in this section to improve the EE while guaranteeing the MCS level
constraint. Different from the previous energy efficient schemes which are only applicable for the Shannon capacity, our proposed scheme determines
the energy efficient transmit power and MCS level according to a practical EE estimation mechanism, which is based on CQI
feedback. Furthermore, we propose a semi-static dual trigger to control the transmit power and MCS level configuration, which is practical in
the HSDPA systems.

Figure 3 shows the operational flowchart of the proposed power control and link adaptation procedure at the Node B. As long as CQI and
acknowledgement/negative acknowledgement(ACK/NACK) information are received by the Node B, Node B can estimate the EE and the required transmit
power for each MCS level based on the estimation mechanism. Then Node B can determine the MCS level and transmit power with maximum EE. After
that, the Node B will determine whether they need to be configured immediately or not, where a semi-static dual trigger
mechanism is employed. If it is triggered, the derived optimal transmit power and corresponding optimal MCS level will be reconfigured. In this way, the scheme is realized in a semi-static manner. There are two benefits here. For one thing, the semi-static feature makes the scheme practical in HSDPA which does not support inner loop power control. For another, the cost of signaling can be reduced significantly through controlling the power reconfiguration cycle length adaptively.

In the following subsections, we will introduce the scheme in details.

\subsection{EE Estimation and Optimal Transmit Power Determination}
We propose the addition of an EE estimation mechanism to the traditional link adaptation operation, whereby it employs the MCS table to estimate the EE and required transmit power for different MCS levels based on CQI feedbacks, and then determines the EE optimal transmit power and MCS level. The MCS table here is defined as the mapping relationship between HS-PDSCH received signal to interference and noise ratio (SINR) threshold and the corresponding feedback CQI index, based on the initial BER target $\mit\Gamma_{\rm tar}$. Each CQI index corresponds to a dedicated MCS level in HSDPA. An example of TABLE G \cite{refer3} is shown in Figure 4.

At first, we need to estimate the transmit power required for different MCS levels. According to \cite{refer13}, the SINR of HS-PDSCH is denoted as
\begin{equation} \label{eq7}
\begin{array}{l}
\rho(P_{HS}) = \displaystyle \frac{{SF \cdot P_{HS} g}}{{(1 - \alpha )I_{{\rm{or}}}  + I_{{\rm{oc}}}  + N_0 W}},
\end{array}
\end{equation}
where $SF$,  $P_{HS}$, $g$, $\alpha$, $I_{\rm{or}}$ and $I_{\rm{oc}}$ denote the spreading factor, HS-PDSCH power,  the instantaneous path gain,
the channel orthogonality factor, the total received power from the serving cell and the inter-cell interference, respectively. As the link level simulation has captured the effect of the inter-code interference, according to (\ref{eq7}), received SINR is proportional to transmit power $P_{HS}$ assuming that the interference is constant. By taking the logarithm on both sides of (\ref{eq7}), we can find that the
difference between two transmit power $P_1$ and $P_2$ is equal to the difference between the two SINR $\rho (P_1)$ and $\rho (P_2)$ derived
from them:
\begin{equation} \label{eq8}
\begin{array}{l}
P_1(\rm dBm) - P_2(\rm dBm) = \rho (P_1)(\rm dB) - \rho (P_2)(\rm dB),
\end{array}
\end{equation}
where transmit power is measured in dBm and SINR is measured in dB.

After replacing the actual SINRs in (\ref{eq8}) By the SINR thresholds in the MCS table, we can utilize the equation to estimate the transmit power required for the MCS levels. In other words, we propose to approximate the difference between the transmit power required for two MCS levels as the difference between the two's SINR thresholds. For example, assume that the current transmit power is $P$ and the feedback CQI index is $i$. For an arbitrary CQI index denoted by $j$, the corresponding SINR threshold is denoted as $\beta_j$ and the MCS level denoted as $\theta_j$. We can estimate the transmit power $P_j$ required for MCS level $\theta_j$ as follows:
\begin{equation} \label{eq9}
\begin{array}{l}
P_j = P + \beta_j - \beta_i + \delta.
\end{array}
\end{equation}
The offset $\delta$ here is to deal with the impact of channel variations which can be determined based on the feedback ACK/NACK information from the user side.

In the simplest case, $\delta$ can be set to zero and (\ref{eq9}) can be rewritten as:
\begin{equation} \label{eq10}
\begin{array}{l}
P_j = P + \beta_j - \beta_i.
\end{array}
\end{equation}
Note that transmit power is measured in dBm and SINR threshold is measured in dB in (\ref{eq9}) and (\ref{eq10}).

One may argue that the adjustment would cause the variation of BER, and then affect the average number of the retransmissions, which may cause the energy wasting. This is not the case. The same BER can be guaranteed for the current and adjusted power level and MCS level, which can be explained as follows. Note that the MCS table at both the BS and the user is based on a fixed BER target. Therefore, it is obvious that the current power level and feedback CQI can guarantee the BER. During the adjustment, to make sure the same BER can be guaranteed, the transmit power and the MCS level are jointly adjusted. That is to say, when the transmit power is decreased, the corresponding MCS level should also be decreased. As the same BER is guaranteed in this way, the same retransmission probability can also be guaranteed, and the average number of the retransmissions will not be affected. In a word, our scheme would work well without affecting the mechanism of the retransmission, which is practical in real systems.

Then the estimation of EE for the MCS level $\theta_j$ is given by:
\begin{equation} \label{eq11}
\begin{array}{l}
\xi_ j = \displaystyle \frac{\tau_j}{{t_{\rm s} \cdot (\frac{{P_j }}{\eta} + P_{{\rm{Dyn}}} + P_{{\rm{Sta}}} )}},
\end{array}
\end{equation}
where $\tau_j$ represents the transport block size of the MCS level $\theta_j$, and $t_{\rm s}$ is equal to two milliseconds and represents the duration of one TTI for HSDPA. Then we compare the estimated EE for each MCS level, determine the optimal CQI index $j^*$ by

\begin{equation} \label{eq12}
\begin{array}{l}
j^* = \mathop {\arg \max } \limits_j \xi _j.
\end{array}
\end{equation}
The corresponding MCS level is denoted as $\theta_{j^*}$ and the required transmit power denoted as $P_{j^*}$.

As the minimum MCS level of the user is $\theta_{\min}$ and the maximum transmit power of the Node B is $P_{\max}$, the constrained optimal MCS level and the optimal transmit power can be given by:
\begin{equation} \label{eq13}
\begin{array}{l} \theta_{\rm opt}  = \min (\max (\theta_{\min } ,{\rm{ }}\theta_{j^*} ),{\rm{ }}\theta_{\max } ), \\
 P_{\rm opt}  = {\rm{ }}\min (\max (P_{\min } ,{\rm{ }}P_{j^*} ),{\rm{ }}P_{\max } ). \\
\end{array}
\end{equation}
The same estimation mechanism above can be employed to determine the corresponding minimum transmit power $P_{\min}$ and the corresponding maximum MCS level $\theta_{\max}$. Correspondingly, the estimated EE for the optimal MCS level and transmit power is denoted as $\xi_{\rm opt}$.

In our proposed algorithm, only the feedback CQI and ACK/NACK information are necessary for Node B to do the EE estimation and energy efficient power determination.

\subsection{Semi-static Power Reconfiguration Trigger}

However, the power configuration cannot be performed instantaneously due to the following two reasons. For one thing, the support for fast AMC and
HARQ functionality in HSDPA does not allow the transmit power change frequently. For another, in order to guarantee the accuracy of the CQI
measurement and user demodulation especially for high order modulation, Node B should inform the user of the transmit power modifications
through the signalling called measurement power offset (MPO) in radio resource control (RRC) layer when the transmit power is reconfigured. If
the configuration performs frequently, the signaling overhead is significant. Therefore, we propose a semi-static trigger mechanism to control
the procedure.

Assume that the EE derived from the last transmission is $\xi$, define relative EE difference $D$ as follows:
\begin{equation} \label{eq14}
\begin{array}{l}
D = \frac{\displaystyle {\xi_{{\rm{opt}}}  - \xi }}{\displaystyle {\xi_{{\rm{opt}}} }}.
\end{array}
\end{equation}
In our proposed scheme, the minimum trigger interval is set to be $\gamma _{\rm prohibit}$, and the maximum trigger interval to be $\gamma _{\rm periodic}$ which satisfies $\gamma _{{\rm{periodic}}}  \gg \gamma_{{\rm{prohibit}}} $. A timer is used to count the time from the last power configuration and the timing is denoted as $t$.

First, if
\begin{equation} \label{eq15}
\left\{\begin{array}{ll}D \ge \Delta\\
t > \gamma _{\rm prohibit}\end{array} \right.
\end{equation}
both are satisfied, the proposed energy efficient power configuration and corresponding MCS reselection process is triggered. This event trigger can guarantee EE gain and also avoid frequent power configuration. On the other hand, if
\begin{equation} \label{eq16}
t > \gamma _{\rm periodic}
\end{equation}
is satisfied, the power configuration process must be triggered regardless of the value of $D$. This periodical trigger ensures that the scheme is always active and gurantees the EE gain. If the power configuration is triggered, the timer must be reset to zero. The whole trigger mechanism above is robust as its parameters can be configured adaptively according to actual systems. It can be implemented practically in HSDPA and signaling overhead can be reduced.

\section{Extension to MIMO System}
As the MIMO technique called D-TxAA can be applied in HSDPA, we propose a modified power control and link adaptation scheme which is applicable to MIMO HSDPA systems in this section.

When MIMO is configured, the Node B will transmit data to the user through either single stream or dual streams in the physical layer. If the former is selected, the proposed scheme in the previous section still works well and the estimated EE is also given by (\ref{eq11}). If the latter is selected, only the EE estimation mechanism in the Node B need to be modified. In this situation, the Node B estimates the sum EE of the two streams instead of a single stream. As transmit power is always shared equally between the two streams, transmit power modifications of the two streams must be the same during the reconfiguration. According to (\ref{eq8}), the corresponding SINR threshold difference between the reconfigured CQI and the previous one is also the same for the two streams. For example, denote the feedback CQI index of the first stream as $i_1$ and the second stream $i_2$. The MCS levels they indicated are $\theta_{i_1}$ and $\theta_{i_2}$ respectively. If the corresponding CQI index for the first stream is adjusted to $j_1$ and that for the second stream is adjusted to $j_2$ when the transmit power is reconfigured, the MCS levels used will be changed into $\theta_{j_1}$ and $\theta_{j_2}$ respectively. Denote the corresponding SINR threshold for CQI index $i_1$, $i_2$, $j_1$ and $j_2$ as $\beta_{i_1}$, $\beta_{i_2}$, $\beta_{j_1}$ and $\beta_{j_2}$, respectively, the following equation must be satisfied:
\begin{equation} \label{eq17}
\begin{array}{l}
\beta_{j_1}-\beta_{i_1} = \beta_{j_2}-\beta_{i_2}.
\end{array}
\end{equation}
The estimation of transmit power required for the new MCS level pair $\theta_{j_1}$ and $\theta_{j_2}$ can be given by
\begin{equation} \label{eq19}
\begin{array}{l}
P_{\rm new}  = P  + 2 \cdot (\beta_{j_1}  - \beta_{i_1} ).
\end{array}
\end{equation}
The estimation of the sum EE can be given by
\begin{equation} \label{eq19}
\begin{array}{l}
\xi_{\rm new}  = \displaystyle \frac{{\tau_{j_1}  + \tau_{j_2} }}{{t_{\rm s} \cdot (\frac{{P_{\rm new} }}{\eta } + P_{{\rm{Dyn}}}  + P_{{\rm{Sta}}} )}}.
\end{array}
\end{equation}
Through comparing the sum EE among all possible MCS level pairs of the two streams, the optimal transmit power and the corresponding MCS level pair for dual streams is selected.

As the mode switching between single stream and dual streams is done at the user side based on maximizing SE, one may argue that the chosen mode may not be the most energy efficient one. Interestingly, as the total power consumption is the same for the two mode according to the power model given by (\ref{eq4}), the choice made at the user side can lead to the most energy efficient mode, which can be explained as follows. Comparing (\ref{eq11}) with (\ref{eq19}), we can know that the denominators of the expressions on the right side are the same, so the value of estimated EE is determined by the numerators. Thus if the sum transport block sizes of the preferred MCS levels for dual stream mode is greater than that for single stream mode, dual stream mode is selected by the user, and vice versa. So the energy efficient criterion for mode selection between single stream and dual streams is the same as maximizing SE criterion.

\section{Simulation Results}
In this section, we evaluate the performance of the proposed algorithm in different scenarios and give some discussions on mode switching between MIMO and SIMO configuration along with our proposed scheme according to HSDPA link level simulation results. A multi-path rayleigh fading channel model and path loss model of PA3 is considered. Bandwidth is 5MHz, and the duration of a subframe is 2ms. The parameters of power model are set as $\eta=0.38$, $P_{\rm cir}=6W$ and $P_{\rm sta}=6W$. The maximum transmit power is set to be 43dBm.

Figure 5 to Figure 7 depict the performance of the proposed semi-static power control method. Proposed energy efficient power control used in
every subframe is viewed as a performance upper bound and the traditional scheme as a baseline where a transmit power of 40.5dBm is configured.
If the energy efficient scheme is used, the transmit power will be configured based on the EE estimation as long as user's feedback is available.
Here parameters of the semi-static trigger are set as $\gamma _{{\rm{prohibit}}}  = 20ms$, $\gamma _{{\rm{periodic}}}  = 200ms$ and $\Delta=20\%$. Figure 5 shows that a considerable EE gain of our proposed semi-static power control scheme can be acquired over the baseline.
Furthermore, the proposed scheme's EE performance is comparable with the upper bound. Figure 6 demonstrates that transmit power reconfiguration frequency is reduced compared with the upper bound algorithm, thus signaling overhead is significantly reduced, due to the proposed dual trigger. The event trigger which sets a threshold for the gap and the periodical trigger also ensures EE gain. Figure 7 also
evaluates the performance of the algorithm under different user speed. We can find that the EE gain would decline with increasing user moving speed, and the reason is explained that when the channel fluctuation becomes faster because of increased moving speed, EE optimal power changes more quickly. However, our proposed power configuration can not track this rapid change due to the semi-static characteristic, so the EE gain decreases, but a considerable EE gain can still be observed at high user speed.

Figure 8 and Figure 9 show the impact of path loss and minimum CQI constrains on EE gain of our proposed scheme. Each minimum CQI constraint corresponds to a minimum MCS constraint. User speed is set as 3 km/h. When the minimum CQI constraints are not so tight, we can see that the EE gain of the proposed algorithm is similar in Figure 8. EE gain decreases when user moves away from Node B and the reason is that the optimal transmit power increases and gradually approaches the transmit power configured in the baseline. From Figure 9, we can also observe that the looser the minimum CQI constrain is, the larger EE gain we can acquire.

Figure 10 gives EE comparison between D-TxAA and SIMO configuration under different transmit power in HSDPA, and Figure 11 illustrates the simulation results for different EE performances between HSDPA-SIMO and HSDPA-MIMO systems by employing our proposed power control method. From Figure 10, we can see that there exists an EE optimal transmit
power for each mode. Another observation is that EE performance of SIMO mode is better than MIMO when transmit power is
not large, and vice versa. The reason is explained as follows. The total power can be divided into three parts: PC power, transmit antenna number related power $P_{\rm Dyn}$, and transmit antenna number independent power $P_{\rm sta}$. When transmit power is large, $P_{{\rm{PC}}}$ dominates the total power (the denominator of the EE) and $P_{{\rm{Dyn}}}$ is negligible. Because the MIMO mode can acquire higher capacity, higher EE is available for this mode in the large transmit power scenario.  When transmit power is low, the ratio of $P_{{\rm{Dyn}}}$ to the total power
increases, and leads to lower EE for MIMO compared with SIMO. Figure 11 provides insights on the impact of the distance on the mode
switching. When the distance between the user and the Node B is getting larger, MIMO is better, and vice versa. This is because in the long
distance scenario, the first part increases and dominates the total power, then more active antenna number is preferred.

From Figure 10 and Figure 11, we can conclude that significant energy saving can be further acquired when adaptive mode switching between SIMO
and MIMO is applied. However, adaptive mode switching may be difficult due to some practical reasons. Firstly, when SIMO mode is configured,
parameters like PCI and CQI for the second stream are not available because the second antenna is switched off to save energy. Thus, how to
estimate the available EE for D-TxAA is a challenge. Secondly, the transmit antenna number information should be informed through the system information,
so the mode switching will impact all users in the cell and bring huge signaling overhead. To sum up, the protocol may need to be redesigned
to utilize the potential EE improvement with mode switching. Nevertheless, the Node B can decide the active antenna number according to the load of the systems,
which should be realized in the network level and is beyond the scope of this paper.

\section{Conclusion}
In this paper, we investigate the impact of transmit power and MCS level configurations on EE in HSDPA and propose an energy efficient semi-static joint power control and link adaptation scheme. We extend the proposed scheme to the MIMO HSDPA scenario. Simulation results prove that the EE gain is significant and the method is robust. Finally, we have a discussion about the potential EE gain of mode switching between SIMO and MIMO configuration along with the practical challenging issues.
\bigskip


\section*{Abbreviations}
HSDPA, high speed downlink packet access; MCS, modulation and coding scheme; MIMO, multiple input multiple output; SIMO, single input multiple output; ICT, Information and communications technology; HS-PDSCH, high speed physical downlink shared channel; AMC, Adaptive Modulation and Coding scheme; HARQ, hybrid Automatic Repeat request; D-TxAA, dual stream transmit adaptive antennas; QoS, Quality of Service; PC, power conversion; BER, bit error rate; SISO, single input single output; CQI, channel quality indicator; PCI, precoding control indicator; RF, Radio Frequency; SE, spectral efficiency; EE, energy efficiency; AWGN, additive white Gaussian noise; SINR, signal to interference and noise ratio; ACK, acknowledgement; NACK , negative acknowledgement; MPO, measurement power offset; RRC, radio resource control.

\section*{Competing interests}
The authors declare that they have no competing interests.

\section*{Acknowledgements}
 This work is supported by Huawei Technologies, Co. Ltd., China.







\newpage
\section*{Figures}

  \subsection*{Figure 1 - SE and EE calculated using shannon capacity formula}
\begin{figure}[h]
\begin{center}
\includegraphics[height = 1.4in] {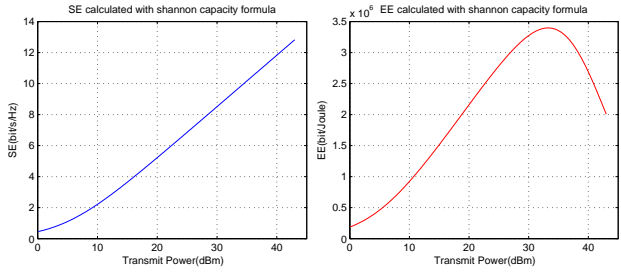}
\end{center}
\caption{SE and EE calculated using shannon capacity formula} \label{fig1}
\end{figure}
  \subsection*{Figure 2 - SE and EE acquired from HSDPA link level simulation}
\begin{figure}[h]
\begin{center}
\includegraphics[height = 1.4in] {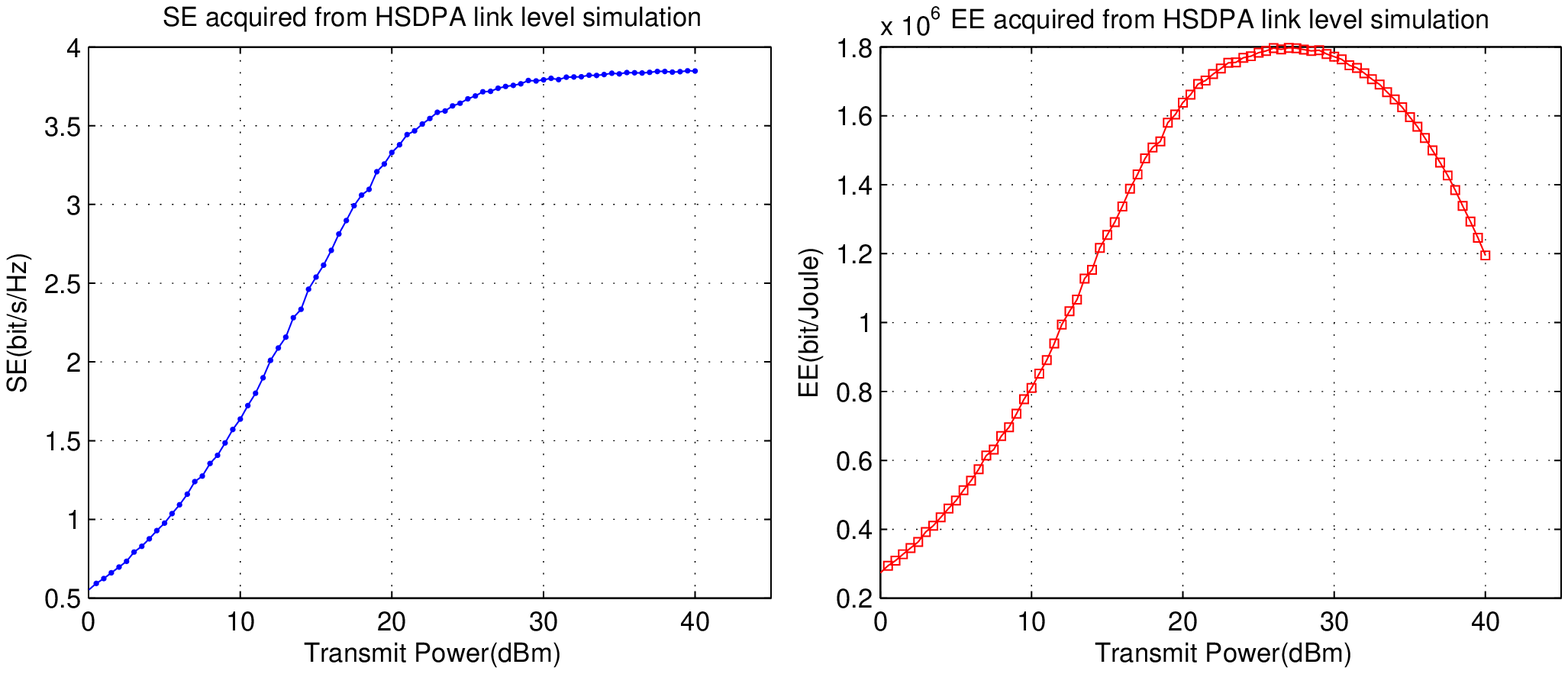}
\end{center}
\caption{SE and EE acquired from HSDPA link level simulation} \label{fig2}
\end{figure}
  \subsection*{Figure 3 - flowchart of the proposed energy efficient power control procedure}
\begin{figure}[h]
\begin{center}
\includegraphics[height = 3in] {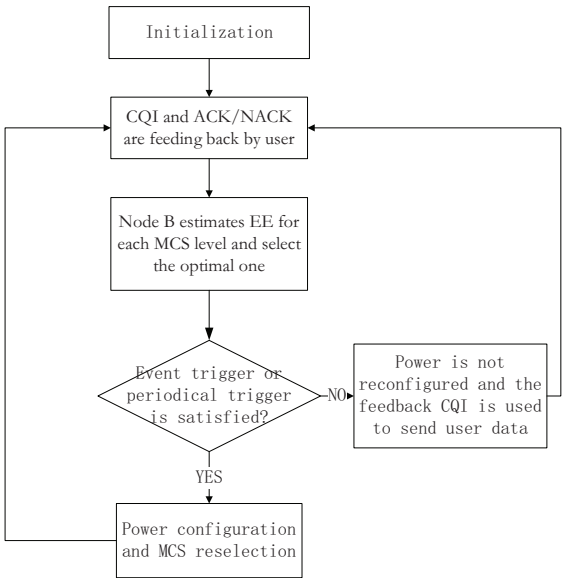}
\end{center}
\caption{flowchart of the proposed energy efficient power control procedure} \label{fig3}
\end{figure}
  \subsection*{Figure 4 - MCS Table for User Category G}
\begin{figure}[h]
\begin{center}
\includegraphics[height = 2.2in] {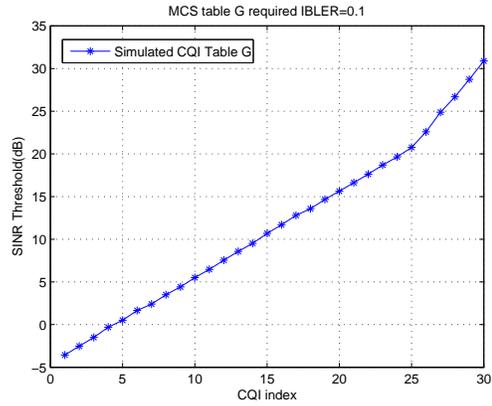}
\end{center}
\caption{MCS Table for User Category G} \label{fig4}
\end{figure}
  \subsection*{Figure 5 - EE comparison of different strategies in 200 subframes}
\begin{figure}[h]
\begin{center}
\includegraphics[height = 2.2in] {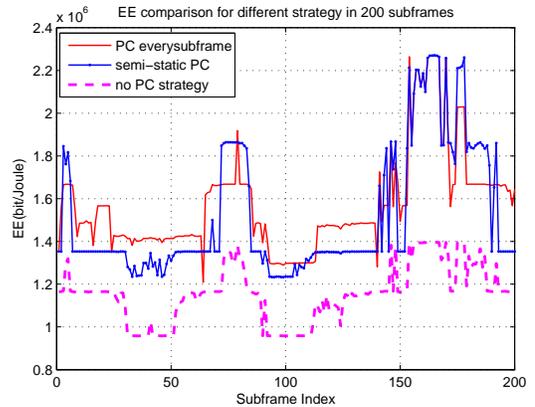}
\end{center}
\caption{EE comparison of different strategies in 200 subframes} \label{fig5}
\end{figure}

\newpage
  \subsection*{Figure 6 - transmit power reconfiguration frequency comparison of different strategies in 200 subframes}
\begin{figure}[h]
\begin{center}
\includegraphics[height = 2.2in] {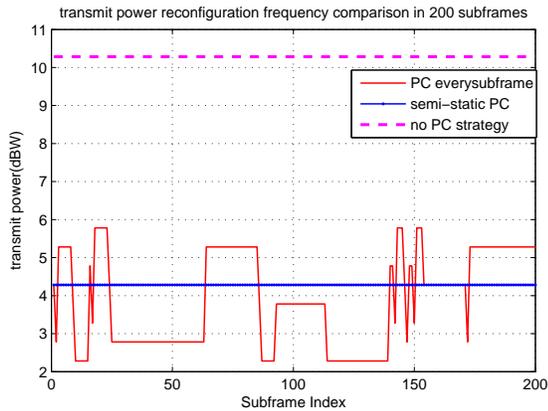}
\end{center}
\caption{transmit power reconfiguration frequency comparison of different strategies in 200 subframes} \label{fig6}
\end{figure}
  \subsection*{Figure 7 - EE comparison of different strategies under different user speed}
\begin{figure}[h]
\begin{center}
\includegraphics[height = 2.2in] {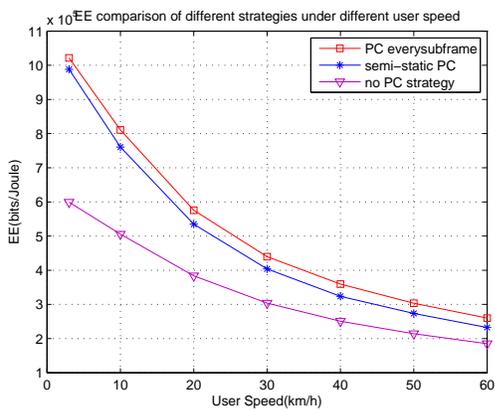}
\end{center}
\caption{EE comparison of different strategies under different user speed} \label{fig7}
\end{figure}

\newpage
  \subsection*{Figure 8 - EE comparison of power control strategies under different user distance from Node B}
\begin{figure}[h]
\begin{center}
\includegraphics[height = 2.2in] {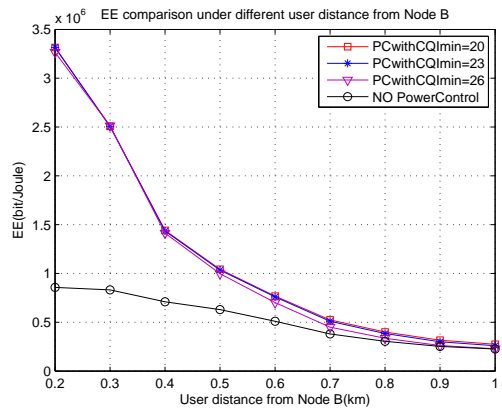}
\end{center}
\caption{EE comparison of power control strategies under different user distance from Node B} \label{fig8}
\end{figure}
  \subsection*{Figure 9 - EE comparison of different strategies under different CQI constraints}
\begin{figure}[h]
\begin{center}
\includegraphics[height = 2.2in] {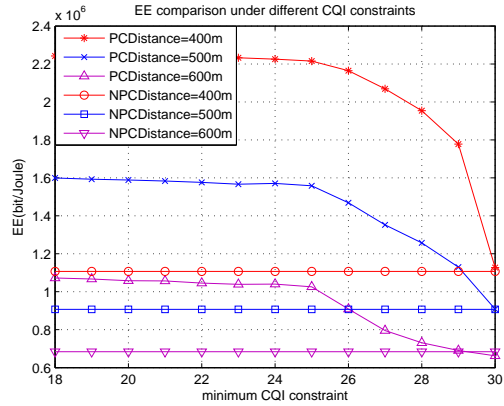}
\end{center}
\caption{EE comparison of different strategies under different CQI constraints} \label{fig9}
\end{figure}

\newpage
  \subsection*{Figure 10 - EE comparison of MIMO and SIMO systems with different transmit power}
\begin{figure}[h]
\begin{center}
\includegraphics[height = 2.2in] {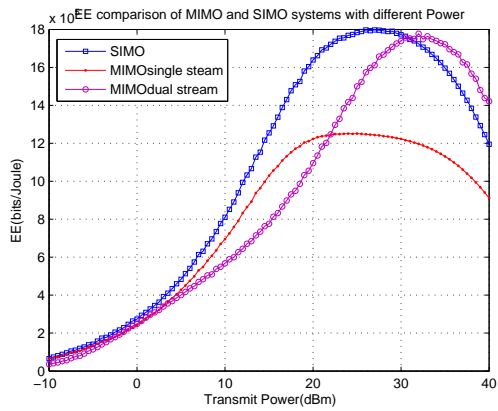}
\end{center}
\caption{EE comparison of MIMO and SIMO systems with different transmit power} \label{fig10}
\end{figure}
  \subsection*{Figure 11 - EE comparison of the proposed strategy for MIMO and SIMO systems in HSDPA with different user pathloss from Node B}
\begin{figure}[h]
\begin{center}
\includegraphics[height = 2.2in] {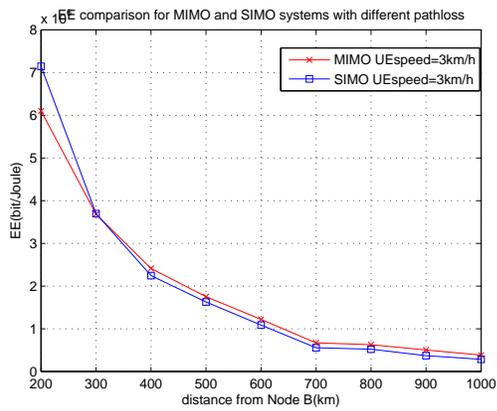}
\end{center}
\caption{EE comparison of the proposed strategy for MIMO and SIMO systems in HSDPA with different user pathloss from Node B} \label{fig11}
\end{figure}

\end{document}